# On Advantages of the Kelvin Mapping in Finite Element Implementations of Deformation Processes

Thomas Nagel[1,2], Uwe-Jens Görke[1], Kevin M. Moerman[2,3], Olaf Kolditz[1,4]

**Abstract** Classical continuum mechanical theories operate on three-dimensional Euclidian space using scalar, vector, and tensor-valued quantities usually up to the order of four. For their numerical treatment, it is common practice to transform the relations into a matrix-vector format. This transformation is usually performed using the so-called Voigt mapping. This mapping does not preserve tensor character leaving significant room for error as stress and strain quantities follow from different mappings and thus have to be treated differently in certain mathematical operations. Despite its conceptual and notational difficulties having been pointed out, the Voigt mapping remains the foundation of most current finite element programmes. An alternative is the so-called Kelvin mapping which has recently gained recognition in studies of theoretical mechanics. This article is concerned with benefits of the Kelvin mapping in numerical modelling tools such as finite element software. The decisive difference to the Voigt mapping is that Kelvin's method preserves tensor character, and thus the numerical matrix notation directly corresponds to the original tensor notation. Further benefits in numerical implementations are that tensor norms are calculated identically without distinguishing stress or strain-type quantities and tensor equations can be directly transformed into matrix equations without additional considerations. The only implementational changes are related to a scalar factor in certain finite element matrices and hence, harvesting the mentioned benefits comes at very little cost.

*Corresponding author. E-mail: thomas.nagel@ufz.de

[1] Helmholtz Centre for Environmental Research – UFZ, Leipzig, Germany.

[2] Trinity College Dublin, Dublin, Ireland.

[3] Media Lab, Massachusetts Institute of Technology, Cambridge, USA.

[4] Technische Universität Dresden, Dresden, Germany.

**Nomenclature**

Throughout the article bold face symbols denote tensors and vectors. Normal face letters represent scalar quantities.

| | |
|---|---|
| $\boldsymbol{\epsilon}$, $\boldsymbol{\epsilon}$, $\underline{\epsilon}$ | Small strain tensor / its Kelvin mapping / its Voigt mapping. |
| $\mathrm{d}\Gamma$ | Area element. |
| $\lambda$ | First Lamé constant. |
| $\mu$ | Second Lamé constant, shear modulus of linear elasticity. |
| $\mathrm{d}\Omega$ | Volume element. |
| $\Psi^i$ | Residual in iteration $i$. |
| $\rho$ | Mass density. |
| $\boldsymbol{\sigma}$, $\boldsymbol{\sigma}$, $\underline{\sigma}$ | Cauchy stress tensor / its Kelvin mapping / its Voigt mapping. |
| $\mathbf{a}\cdot\mathbf{b}$ | Dot product of $\mathbf{a}$ and $\mathbf{b}$. |
| $\mathbf{A}:\mathbf{B}$ | Double contraction of $\mathbf{A}$ and $\mathbf{B}$. |
| $\mathbf{a}\otimes\mathbf{b}$ | Dyadic product of $\mathbf{a}$ and $\mathbf{b}$. |
| $\mathbf{A}\underline{\odot}\mathbf{B}$ | $\frac{1}{2}\left[(\mathbf{A}\otimes\mathbf{B})^{\overset{23}{\mathrm{T}}}+(\mathbf{A}\otimes\mathbf{B}^{\mathrm{T}})^{\overset{24}{\mathrm{T}}}\right]$ |
| $\mathbf{A}\odot\mathbf{B}$ | $\frac{1}{2}\left[(\mathbf{A}\otimes\mathbf{B}^{\mathrm{T}})^{\overset{23}{\mathrm{T}}}+(\mathbf{A}\otimes\mathbf{B})^{\overset{24}{\mathrm{T}}}\right]$ |
| div | Divergence operator. |
| grad | Gradient operator. |
| $(\bullet)^{\mathrm{D}}$ | Deviatoric part of a tensor. |
| $(\bullet)^{\mathrm{T}}$, $(\bullet)^{\overset{ab}{\mathrm{T}}}$ | Transpose operator; transposition of a$^{\text{th}}$ and b$^{\text{th}}$ base vector. |
| $\mathbf{b}$ | External body force. |
| $\mathscr{C}$, $\boldsymbol{C}$, $\underline{\underline{C}}$ | Tangent moduli (fourth order tensor) / its Kelvin mapping / its Voigt mapping. |
| $\mathbf{E}$ | Green Lagrange strain tensor. |
| $e$ | Linear volume strain. |
| $\mathbf{F}$ | Deformation gradient. |
| $F$ | Yield function. |
| $G$ | Plastic potential. |
| $J$ | Volume ratio of material volume elements in the current and the reference configuration. |
| $K$ | Bulk modulus of linear elasticity. |
| $\mathbf{n}$ | Outward unit normal vector. |
| $N_a$, $\mathbf{N}$ | Nodal shape function, element matrix of nodal shape functions. |
| $\mathscr{P}^{\mathrm{S}}$, $\mathscr{P}^{\mathrm{D}}$ | Spherical and deviatoric projection tensors. |
| $p$ | Hydrostatic pressure. |
| $\mathbf{S}$ | Second Piola Kirchhoff stress tensor. |
| $\mathbf{t}$ | Surface traction. |
| $\mathbf{u}$ | Displacement vector. |

# 1 Introduction

Classical continuum mechanical theories operate on three-dimensional Euclidian space using scalar, vector, and tensor-valued quantities usually up to the order of four. For example, the generalised Hooke's law in linear elasticity reads in symbolic tensor notation

$$\boldsymbol{\sigma} = \mathscr{C} : \boldsymbol{\epsilon} \qquad (1)$$

Eq. (1) cannot be used directly in numerical software the implementation of which is based on matrix-vector algebra. It is hence common practice to transform Eq. (1) and similar formulations into a matrix-vector format in order to make use of powerful linear algebraic manipulation tools. This transformation is usually performed using the so-called Voigt mapping (Voigt 1966). It is introduced as a simple replacement of fourth-order tensors having the necessary symmetries by 6×6 matrices and symmetric second order tensors by 6×1 vectors. This mapping, however, is performed differently for stress- and strain-type quantities. Based on the tensor coordinate matrices in a particular basis the transformation is performed as follows:

$$\sigma_{ij} \to \underline{\sigma} = [\sigma_{11}\ \sigma_{22}\ \sigma_{33}\ \sigma_{12}\ \sigma_{23}\ \sigma_{13}\ ]^{\mathrm{T}} \quad (2)$$

$$\epsilon_{ij} \to \underline{\epsilon} = [\epsilon_{11}\ \epsilon_{22}\ \epsilon_{33}\ 2\epsilon_{12}\ 2\epsilon_{23}\ 2\epsilon_{13}\ ]^{\mathrm{T}} \quad (3)$$

The use of engineering shear strains $\gamma_{ij} = 2\epsilon_{ij}$ is apparent and has been introduced to maintain the following useful identities

$$\boldsymbol{\sigma} : \boldsymbol{\epsilon} = \sigma_{ij}\epsilon_{ij} = \underline{\sigma} \cdot \underline{\epsilon} \quad (4)$$

Furthermore, the constitutive matrix $\underline{\underline{C}}$ directly contains the entries of the stiffness tensor $\mathscr{C}$. The coordinates of fourth-order tensors are "manually" rearranged into 6×6 matrices so as to ensure that the resulting matrix-vector operations yield results that are mathematically equivalent to the original tensor operation, for example

$$\underline{\sigma} = \underline{\underline{C}}\,\underline{\epsilon} \quad (5)$$

corresponding to Eq. (1). This procedure provides significant room for error that can in principle be avoided. As apparent in Eqs. (2) and (3), stress and strain quantities follow from different mappings and thus have to be treated differently in certain mathematical operations, e.g. the calculation of tensor norms (as will be shown in Sections 2.2 and 3). The reason for the separate numerical treatment of the same mathematical concept is that the Voigt mapping does not preserve tensor character. The same (mathematical) function therefore has to be implemented multiple times into software when used with different quantities. For complex models, it becomes increasingly, and unnecessarily, difficult to keep track of the necessary distinctions. The Voigt mapping has remained a standard method in some fields even today, despite the conceptual and notational difficulties having been pointed out (Mehrabadi and Cowin 1990). It is also the foundation of most current finite element programmes (Bonet and Wood 1997; Bathe 2001; Zienkiewicz et al 2006; Wriggers 2008; Hughes 2012) .

An alternative to the Voigt mapping can be found in the so-called Kelvin mapping. It is based on an article by Lord Kelvin (Thomson 1856) on elasticity theory where he represented stress and strain not as second-order tensors in 3D space but as 6D vectors. An interesting historical review, a "translation" into modern concepts and a hypothetical continuation of Kelvin's article – which did not receive much attention until more than century later – can be found in Helbig (2013). The recently increasing interest and recognition of the Kelvin mapping has clustered around studies of theoretical mechanics (see Section 2.1). This article, however, is concerned with a useful "side-effect" of the approach. The aim here is to highlight benefits of the Kelvin mapping in numerical modelling tools such as finite element software. Because in Kelvin's approach, stresses and strains are represented in a 6D vector space, Hooke's law can be written with the help of a second-order 6D stiffness tensor as

$$\boldsymbol{\sigma} = \mathbf{C}\boldsymbol{\epsilon} \tag{6}$$

While Eqs. (5) and (6) are structurally similar, the decisive difference is that Kelvin's method preserves tensor character, and thus the numerical matrix notation directly corresponds to the original tensor notation. This means that implemented equations and equations derived on paper are practically identical without individual mappings for stress/strain quantities, deviatoric and spherical components etc. This becomes increasingly valuable for complex mechanical models and in the context of a Newton-Raphson solution scheme, where many derivations of tensor-valued functions have to be performed.

In the context of our work, such models have been implemented into the scientific open-source platform OpenGeoSys (Kolditz et al 2012) to simulate the mechanical behaviour of geotechnical materials such as salt rock (Heusermann et al 2003; Minkley and Mühlbauer 2007) for use in subsurface energy storage (Bauer et al 2013; Li et al 2015; Ma et al 2015).

## 2 The Kelvin mapping

### 2.1 Applications in elasticity theory – a short review

The ideas of Kelvin have been rediscovered in the context of anisotropic elasticity, see Mehrabadi and Cowin (1990); Kowalczyk-Gajewska and Ostrowska-Maciejewska (2014), and put into the context of modern tensor algebra by several authors (Mehrabadi and Cowin 1990; Kowalczyk-Gajewska and Ostrowska-Maciejewska 2014; Moakher 2008). Major topics of interest in which the concept has been used are: the use of six eigenstiffnesses and orthogonal eigenstates for a better understanding of material behaviour (Rychlewski 1984; Annin and Ostrosablin 2008); different aspects of a spectral decompositon of the stiffness tensor (Theocaris and Philippidis 1991; Theocaris 2000; Bolcu et al 2010); the investigation of material symmetries and preferred deformation modes of anisotropic media, e.g. composite materials (Mehrabadi and Cowin 1990; Bóna et al 2007) including the relationship to fabric tensors (Moesen et al 2012) and deformation-induced anisotropy (Cowin 2011); the transformation of the properties of one anisotropic medium to the closest effective medium from a differing symmetry group (Norris 2006; Diner et al 2011; Kochetov and Slawinski 2009; Moakher and Norris 2006); wave attenuation and elastic constant inversion from wave traveltime data (Carcione et al 1998; Dellinger et al 1998). The inversion of Hooke's law in the case of incompressible or slightly compressible materials was studied by Itskov and Aksel (2002), while the use of the spectral decomposition of the stiffness tensor in a constitutive formulation for finite hyper-elasticity in a finite element context was described in Dłuzewski and Rodzik (1998). For further examples of the application of the Kelvin mapping in modern mechanics, see the references in the works cited above, especially Helbig (2013). In the sequel, the focus will be on benefits of the Kelvin mapping in finite element schemes that are a useful "side effect" of the concept.

2.2 Spectral decompositions and the Kelvin basis

In this section, the spectral decomposition of fourth order tensors is very briefly reviewed as it provides a natural access to the Kelvin mapping in finite element implementations.

The eigenvalue problem of a second-order tensor $\mathbf{A}$ can be written as

$$\mathbf{A}\mathbf{n} = \lambda \mathbf{n} \quad \text{with} \quad \mathbf{n} \neq \mathbf{0} \tag{7}$$

Using $\mathbf{I} = \mathbf{n}_i \otimes \mathbf{n}_i$ with $\mathbf{n}_i \cdot \mathbf{n}_j = \delta_{ij}$ yields the spectral decomposition of a second order tensor:

$$\mathbf{A} = \mathbf{A}\mathbf{I} = \mathbf{A}(\mathbf{n}_i \otimes \mathbf{n}_i) = (\mathbf{A}\mathbf{n}_i) \otimes \mathbf{n}_i \tag{8}$$

$$= \sum_{i=1}^{3} \lambda_{(i)} \mathbf{n}_{(i)} \otimes \mathbf{n}_{(i)} = \lambda_i \mathbf{N}_i \tag{9}$$

with the three eigenvalues $\lambda_i$ and the three second-order eigenprojections $\mathbf{N}_i$ composed of the eigenvectors $\mathbf{n}_i$. Regarding notation, an index occuring twice implies summation following Einstein's summation convention. If an index is written in parentheses, e.g. $(i)$, this convention is suppressed.

A similar problem can be posed for super-symmetric fourth order tensors (Itskov 2009):

$$\mathscr{A} : \mathbf{M} = \Lambda \mathbf{M} \quad \text{with} \quad \mathbf{M} \neq \mathbf{0} \tag{10}$$

Noting that $\mathbf{M}_i : \mathbf{M}_j = \delta_{ij}$, the fourth-order super-symmetric identity tensor $\mathscr{I}^\text{s}$ with the property $\mathscr{I}^\text{s} : \mathbf{A} = \text{sym}\,\mathbf{A}$ can be written in terms of the tensors $\mathbf{M}$ (Kowalczyk-Gajewska and Ostrowska-Maciejewska 2014):

$$\mathscr{I}^\text{s} = \mathbf{I} \underline{\odot} \mathbf{I} = \sum_{i=1}^{6} \mathbf{M}_{(i)} \otimes \mathbf{M}_{(i)} \tag{11}$$

Using this identity yields the spectral decomposition of a fourth order tensor:

$$\mathscr{A} = \mathscr{A} : \mathscr{I}^\text{s} = (\mathscr{A} : \mathbf{M}_i) \otimes \mathbf{M}_i \tag{12}$$

$$= \sum_{i=1}^{6} \Lambda_{(i)} \mathbf{M}_{(i)} \otimes \mathbf{M}_{(i)} = \Lambda_i \mathscr{M}_i \tag{13}$$

with the six eigenvalues $\Lambda_i$ and the six fourth-order eigenprojections $\mathscr{M}_i$ composed of the second-order eigentensors $\mathbf{M}_i$.

Comparison of Eqs. (11) and (13) shows that $\mathscr{I}^\text{s}$ has the eigenvalue $\Lambda = 1$ of multiplicity 6. Choosing a Cartesian basis $\{\mathbf{e}_i\}$ and noticing $\mathbf{I} = \mathbf{e}_i \otimes \mathbf{e}_i$ allows the representation

$$\mathscr{I}^\text{s} = \mathbf{I} \underline{\odot} \mathbf{I} = \frac{1}{2} \left[ \mathbf{e}_i \otimes \mathbf{e}_j \otimes \mathbf{e}_i \otimes \mathbf{e}_j + \mathbf{e}_i \otimes \mathbf{e}_j \otimes \mathbf{e}_j \otimes \mathbf{e}_i \right] \tag{14}$$

By using Eq. (11), one finds the six eigentensors of $\mathscr{I}^\text{s}$ (Itskov 2009):

$$\mathbf{M}_1 = \mathbf{e}_1 \otimes \mathbf{e}_1 \qquad\qquad \mathbf{M}_4 = \frac{1}{\sqrt{2}}(\mathbf{e}_1 \otimes \mathbf{e}_2 + \mathbf{e}_2 \otimes \mathbf{e}_1)$$

$$\mathbf{M}_2 = \mathbf{e}_2 \otimes \mathbf{e}_2 \qquad\qquad \mathbf{M}_5 = \frac{1}{\sqrt{2}}(\mathbf{e}_2 \otimes \mathbf{e}_3 + \mathbf{e}_3 \otimes \mathbf{e}_2) \tag{15}$$

$$\mathbf{M}_3 = \mathbf{e}_3 \otimes \mathbf{e}_3 \qquad \mathbf{M}_6 = \frac{1}{\sqrt{2}}(\mathbf{e}_1 \otimes \mathbf{e}_3 + \mathbf{e}_3 \otimes \mathbf{e}_1)$$

These eigentensors can be viewed as the basis of the Kelvin mapping. Instead of simply reordering tensor coordinates, the Kelvin mapping proceeds from the introduction of a new 6D basis $\{\mathbf{E}_I\}$ based on the original 3D basis $\{\mathbf{e}_i\}$ (compare Mehrabadi and Cowin (1990) and the appendix in Cowin and Doty (2007)) by setting

$$\mathbf{E}_I = \mathbf{M}_I(\mathscr{I}^\mathrm{s}) \; \forall I = 1, \ldots, 6 \tag{16}$$

In other words, this basis is identical to the eigentensors of the symmetry pojection tensor $\mathscr{I}^\mathrm{s}$, compare Eq. (15).

Thus, exemplary tensors with the necessary symmetries can equivalently be written in the various bases

$$\mathbf{A} = A_{ij}\mathbf{e}_i \otimes \mathbf{e}_j = A_I \mathbf{E}_I \qquad \text{with} \quad A_I = \mathbf{A} : \mathbf{E}_I \tag{17}$$

$$\mathscr{A} = A_{ijkl}\mathbf{e}_i \otimes \mathbf{e}_j \otimes \mathbf{e}_k \otimes \mathbf{e}_l = A_{IJ} \mathbf{E}_I \otimes \mathbf{E}_J \qquad \text{with} \quad A_{IJ} = \mathbf{E}_I : \mathscr{A} : \mathbf{E}_J \tag{18}$$

One can see that, similar to the Voigt mapping, the coordinates of second and fourth order tensors can now be represented as 6-dimensional vectors and matrices. However, the tensor character of all quantities is still preserved. Note further, that the coordinates of the Kelvin mapping of a fourth-order tensor $\mathbf{A} \otimes \mathbf{A}$ simply follow from the coordinate matrix of the dyadic product of the Kelvin mapped vectors. Thus, the same notation can be employed in both cases.

For numerical implementation, the coordinates of the Kelvin-mapped stress and strain tensors can now be used

$$\sigma_{ij} \to \boldsymbol{\sigma} = \begin{bmatrix} \sigma_{11} & \sigma_{22} & \sigma_{33} & \sqrt{2}\sigma_{12} & \sqrt{2}\sigma_{23} & \sqrt{2}\sigma_{13} \end{bmatrix}^\mathrm{T} \tag{19}$$

$$\epsilon_{ij} \to \boldsymbol{\epsilon} = \begin{bmatrix} \epsilon_{11} & \epsilon_{22} & \epsilon_{33} & \sqrt{2}\epsilon_{12} & \sqrt{2}\epsilon_{23} & \sqrt{2}\epsilon_{13} \end{bmatrix}^\mathrm{T} \tag{20}$$

which have the same structure regardless of whether they are stresses or strains. This has important consequences for example for the calculation of tensor norms.

2.3 Tensor norms

Since the Voigt mapping is a simple reorganisation from tensor coordinates into matrix/vector entries, the tensor character of the individual quantities is lost and tensor norms are not maintained. On the other hand, tensor character and hence norms are preserved for the Kelvin mapping.

| original tensor notation | Kelvin mapping | Voigt mapping |
| --- | --- | --- |
| $\boldsymbol{\sigma} : \boldsymbol{\epsilon}$ | $\boldsymbol{\sigma} \cdot \boldsymbol{\epsilon}$ | $\underline{\sigma} \cdot \underline{\epsilon}$ |
| $\boldsymbol{\sigma} : \boldsymbol{\sigma}$ | $\boldsymbol{\sigma} \cdot \boldsymbol{\sigma}$ | $\underline{\sigma} \cdot \underline{\underline{P_2}} \, \underline{\sigma}$ |
| $\boldsymbol{\epsilon} : \boldsymbol{\epsilon}$ | $\boldsymbol{\epsilon} \cdot \boldsymbol{\epsilon}$ | $\underline{\epsilon} \cdot \underline{\underline{P_1}} \, \underline{\epsilon}$ |

**Table 1** Comparison of different tensor norms.

As illustrated in Table 1, in order to compute certain tensor norms, which are often used to define stress- or strain-dependent quantities, different functions need to be implemented for strain- and stress-type quantities when using the Voigt mapping (or one function called with different projection matrices or forefactors). For notational clarification, the following 6×6 projection matrices are defined and used in Table 1:

$$\underline{\underline{P}}_2 = \begin{pmatrix} \underline{\underline{I}} & \underline{\underline{0}} \\ \underline{\underline{0}} & 2\underline{\underline{I}} \end{pmatrix} \quad \text{and} \quad \underline{\underline{P}}_{\frac{1}{2}} = \begin{pmatrix} \underline{\underline{I}} & \underline{\underline{0}} \\ \underline{\underline{0}} & \frac{1}{2}\underline{\underline{I}} \end{pmatrix} \tag{21}$$

No projection matrices equivalent to those defined in (21) are needed when the Kelvin mapping is used; all tensor norms can be directly computed as defined in the original tensor notation. When using the Voigt mapping, one has to continuously keep track of the implementational differences when performing further derivations and calculations, as the following section shows.

## 3 Application examples

3.1 Scalar material parameters

A decomposition of the Cauchy stress tensor into spherical and deviatoric parts yields additional physical insight by separating the hydrostatic from the deviatoric stress state:

$$\boldsymbol{\sigma} = -p\,\mathbf{I} + \boldsymbol{\sigma}^\mathrm{D} \quad \text{with} \quad p = -\frac{1}{3}\boldsymbol{\sigma} : \mathbf{I} \tag{22}$$

Similarly, the small strain tensor can be decomposed into the volumetric and the deviatoric strain:

$$\boldsymbol{\epsilon} = \frac{1}{3} e\,\mathbf{I} + \boldsymbol{\epsilon}^\mathrm{D} \quad \text{with} \quad e = \boldsymbol{\epsilon} : \mathbf{I} \tag{23}$$

The fourth order stiffness tensor is defined via the following relationship:

$$\mathrm{d}\boldsymbol{\sigma} = \mathscr{C} : \mathrm{d}\boldsymbol{\epsilon} \tag{24}$$

Linear elasticity can be parameterised by the Lamé coefficients $\lambda$ and $\mu$. With the common engineering constants shear modulus $G = \mu$ and bulk modulus $K = \lambda + 2/3\mu$ we find the elasticity modulus and the stress-strain relationship with a stress-free reference state as

$$\mathscr{C} = 3K\mathscr{P}^\mathrm{S} + 2G(\mathscr{I}^\mathrm{s} - \mathscr{P}^\mathrm{S}) \tag{25}$$

$$\boldsymbol{\sigma} = K e\,\mathbf{I} + 2G\boldsymbol{\epsilon}^\mathrm{D} \tag{26}$$

where the fourth order projection tensors have been defined in A. A simple extension of this material model to finite deformation measures for the use under large rotations is the St. Venant-Kirchhoff model expressed in terms of the second Piola-Kirchhoff stress and the Green-Lagrange strain tensor:

$$\mathscr{C}_\mathrm{m} = 3K\mathscr{P}^\mathrm{S} + 2G(\mathscr{I}^\mathrm{s} - \mathscr{P}^\mathrm{S}) \tag{27}$$

$$\mathbf{S} = K(\mathbf{E} : \mathbf{I})\mathbf{I} + 2G\,\mathbf{E}^\mathrm{D} \tag{28}$$

| original tensor notation | Kelvin mapping | Voigt mapping |
|---|---|---|
| $\mathrm{d}\boldsymbol{\sigma} = \mathscr{C} : \mathrm{d}\boldsymbol{\epsilon}$ | $\mathrm{d}\boldsymbol{\sigma} = \boldsymbol{C}\mathrm{d}\boldsymbol{\epsilon}$ | $\mathrm{d}\underline{\sigma} = \underline{\underline{C}}\,\mathrm{d}\underline{\epsilon}$ |
| $\mathrm{d}\boldsymbol{\sigma}^\mathrm{D} = \mathscr{C} : \mathrm{d}\boldsymbol{\epsilon}^\mathrm{D}$ | $\mathrm{d}\boldsymbol{\sigma}^\mathrm{D} = \boldsymbol{C}\mathrm{d}\boldsymbol{\epsilon}^\mathrm{D}$ | $\mathrm{d}\underline{\sigma}^\mathrm{D} = \underline{\underline{C}}\,\mathrm{d}\underline{\epsilon}^\mathrm{D}$ |
| $\mathrm{d}\boldsymbol{\sigma}^\mathrm{D} = 2G\mathrm{d}\boldsymbol{\epsilon}^\mathrm{D}$ | $\mathrm{d}\boldsymbol{\sigma}^\mathrm{D} = 2G\mathrm{d}\boldsymbol{\epsilon}^\mathrm{D}$ | $\mathrm{d}\underline{\sigma}^\mathrm{D} = G\mathrm{d}\underline{\epsilon}^\mathrm{D} \neq 2G\mathrm{d}\underline{\epsilon}^\mathrm{D}$ |

**Table 2** Comparison of matrix-based and scalar material parameters. Note, that the Voigt and Kelvin matrices, $\underline{\underline{C}}$ and $\boldsymbol{C}$, differ in some entries by a scalar factor (see, e.g., Cowin and Doty (2007) and Section 4.3). The relations for the Saint Venant-Kirchhoff model are completely analogous.

The use of the Voigt mapping can be a pitfall in FE implementations when switching from matrix-based to scalar material parameters, as highlighted in Table 2.

While in the case of matrix-based material parameters (if the matrices $\boldsymbol{C}$ and $\underline{\underline{C}}$ are properly defined), the implementation follows directly the tensor notation, this only holds true for the Kelvin mapping once scalar material parameters are used.

3.2 Flow rules and yield functions

Consider a general yield function $F$ or plastic potential $G$ of an elastoplastic material expressed in terms of the invariants

$$I_1 = \boldsymbol{\sigma} : \mathbf{I} \qquad J_2 = \frac{1}{2}\boldsymbol{\sigma}^\mathrm{D} : \boldsymbol{\sigma}^\mathrm{D} \qquad J_3 = \det \boldsymbol{\sigma}^\mathrm{D} \tag{29}$$

A stress integration algorithm (compare Section 4.4) usually requires a plastic flow rule, such as

$$\dot{\boldsymbol{\epsilon}}_\mathrm{p} = \lambda \frac{\partial G}{\partial \boldsymbol{\sigma}} \tag{30}$$

in the geometrically linear case. Additionally, the yield condition $F = 0$ is often added as a constraint to ensure that a calculated stress state remains on the yield surface as states with $F > 0$ are usually not admissible. Thus, for a Newton-Raphson iteration to integrate the constitutive model, the derivatives

$$\frac{\partial F}{\partial \boldsymbol{\sigma}} \quad \text{and} \quad \frac{\partial^2 G}{\partial \boldsymbol{\sigma}^2} \tag{31}$$

are needed, among others. Depending on the mathematical representation of $F$ and $G$, these functions can be quite complex so it is desirable to be able to directly transform the tensorial formulation into its numerical equivalent without having to keep track of additional projections or different implementations.

As a simple but illustrative example, consider the following expansion:

$$\frac{\partial G}{\partial \boldsymbol{\sigma}} = \frac{\partial G}{\partial I_1}\frac{\partial I_1}{\partial \boldsymbol{\sigma}} + \frac{\partial G}{\partial J_2}\frac{\partial J_2}{\partial \boldsymbol{\sigma}^\mathrm{D}} : \mathscr{P}^\mathrm{D} + \frac{\partial G}{\partial J_3}\frac{\partial J_3}{\partial \boldsymbol{\sigma}^\mathrm{D}} : \mathscr{P}^\mathrm{D} \tag{32}$$

where $\mathscr{P}^\mathrm{D}$ has been defined in A.

The calculation of $J_2$ requires different routines in the case of Voigt and Kelvin mappings as shown in Section 2.3. Moreover, this difference has to be kept in mind when transforming the partial derivative $\partial J_2/\partial \boldsymbol{\sigma}^\mathrm{D}$ from tensor notation to its numerical implementation. The differences are highlighted in Table 3.

| original tensor notation | Kelvin mapping | Voigt mapping |
| --- | --- | --- |
| $J_2 = \dfrac{1}{2}\boldsymbol{\sigma}^{\mathrm{D}} : \boldsymbol{\sigma}^{\mathrm{D}}$ | $J_2 = \dfrac{1}{2}\boldsymbol{\sigma}^{\mathrm{D}} \cdot \boldsymbol{\sigma}^{\mathrm{D}}$ | $J_2 = \dfrac{1}{2}\underline{\sigma}^{\mathrm{D}} \cdot \underline{\underline{P_2}}\,\underline{\sigma}^{\mathrm{D}}$ |
| $\dfrac{\partial J_2}{\partial \boldsymbol{\sigma}^{\mathrm{D}}} = \boldsymbol{\sigma}^{\mathrm{D}}$ | $\dfrac{\partial J_2}{\partial \boldsymbol{\sigma}^{\mathrm{D}}} = \boldsymbol{\sigma}^{\mathrm{D}}$ | $\dfrac{\partial J_2}{\partial \underline{\sigma}^{\mathrm{D}}} = \underline{\underline{P_2}}\,\underline{\sigma}^{\mathrm{D}}$ |

**Table 3** Comparison of Kelvin and Voigt mapping in relation to the second invariant of the deviatoric stress tensor.

It can be seen again, that the implementation of routines that calculate $J_2$ and its derivative follow directly from the tensor equation in the case of the Kelvin mapping, while the use of the projection matrix to calculate $J_2$, in case of the Voigt mapping also has to be taken into account when calculating its derivative.

## 4 Notes on the implementation

To show that the benefits of the Kelvin mapping can be harnessed at very little cost, this section briefly illustrates implementational differences in a finite element realisation. These changes can be incorporated into the "core" of the FE software, as illustrated here. There is, however, also the possibility to use Kelvin mapping in software without complete access to the sources. If, for example, one wants to implement a user-defined material (UMAT in Abaqus®, HYPELA2 in MSC Marc®), one can simply re-map the incoming quantities and transform the results back into Voigt mapping. This can be worthwhile if the stress integration of the constitutive model is complex enough to warrant the additional effort.

In the sequel, both the equilibrium iteration scheme of the global displacement based formulation of the solid momentum balance as well as the local stress update procedures are based on a consistent linearisation in the context of a Newton-Raphson procedure in order to efficiently couple both implementation levels.

Finite deformations will be considered. In order to keep the presentation short, basic knowledge of the associated concepts is assumed. Without further introduction, consider given a current configuration and a reference configuration as mappings of a physical body into three dimensional space at current and initial time, $t$ and $t_0$, respectively. They are characterised by the covariant bases $\{\mathbf{g}_i\}$ and $\{\mathbf{G}_I\}$, respectively, as well as their corresponding contravariant counterparts $\{\mathbf{g}^i\}$ and $\{\mathbf{G}^I\}$. The deformation gradient is defined as a linear mapping between line elements from the reference to the current configuration:

$$\mathrm{d}\mathbf{x} = \mathbf{F}\,\mathrm{d}\mathbf{X} \quad \text{with} \quad \mathbf{F} = \frac{\partial x^i}{\partial X^J}\mathbf{g}_i \otimes \mathbf{G}^J \tag{33}$$

For more details on this matter consult, e.g., Itskov (2009); Haupt (2002).

### 4.1 The weak form in a material setting

The local form of the quasistatic equilibrium conditions derived from the momentum balance considering body forces $\mathbf{f}$ reads in the current configuration

$$\mathrm{div}\,\boldsymbol{\sigma} + \rho\mathbf{f} = \mathbf{0} \tag{34}$$

and is the basis to determine the displacement field **u** in a domain $\Omega$. A kinematically compatible test function **v** is chosen with $\mathbf{v} = \mathbf{0}\ \forall \mathbf{x} \in \partial \Omega_\mathbf{u}$, i.e. **v** vanishes on the Dirichlet boundary $\partial \Omega_\mathbf{u}$. Transformation of Eq. (34) into the weak form

$$\int_\Omega \boldsymbol{\sigma} : \operatorname{grad} \mathbf{v}\, \mathrm{d}\Omega = \int_\Omega \rho \mathbf{f} \cdot \mathbf{v}\, \mathrm{d}\Omega + \int_{\partial \Omega_\mathbf{t}} \bar{\mathbf{t}} \cdot \mathbf{v}\, \mathrm{d}\Gamma \tag{35}$$

naturally yields the prescribed surface traction vector $\bar{\mathbf{t}}$ on the Neumann boundary $\partial \Omega_\mathbf{t}$ as a result of partial integration, where $\partial \Omega_\mathbf{u} \cup \partial \Omega_\mathbf{t} = \partial \Omega$ and $\partial \Omega_\mathbf{u} \cap \partial \Omega_\mathbf{t} = \emptyset$.

The geometrical configuration of the integration domain in Eq. (35) is unknown and an outcome of the deformation problem to be solved. In that sense, Cauchy stresses are not additive, i.e. $\boldsymbol{\sigma}^{t+\Delta t} \neq \boldsymbol{\sigma}^t + \Delta \boldsymbol{\sigma}$. Thus, Eq. (35) needs to be transformed to a suitable reference configuration on which second Piola-Kirchhoff stresses will be defined that in turn can be decomposed additively. The two most common choices for this reference configuration are the initial (undeformed) configuration at time $t = 0$ leading to a Total Lagrangian (TL) formulation, and the configuration determined in the iteration leading to the Updated Lagrangian (UL) formulation (Bathe 2001; Zienkiewicz et al 2006). Here, a TL formulation is chosen. Similar considerations apply to the UL formulation.

The weak form of Eq. (35) pulled back into the reference configuration reads (compare B)

$$\int_{\Omega^0} \mathbf{S} : \mathbf{E}(\mathbf{U};\mathbf{V})\, \mathrm{d}\Omega^0 = \int_{\Omega^0} \rho_0 \mathbf{F} \cdot \mathbf{V}\, \mathrm{d}\Omega^0 + \int_{\partial \Omega_\mathbf{T}^0} \bar{\mathbf{T}} \cdot \mathbf{V}\, \mathrm{d}\Gamma^0 \tag{36}$$

4.2 Linearisation of the weak form

Eq. (36) is to be solved for a time increment $\Delta t$ to obtain the solution at the next time step $t + \Delta t$ based on the known solution at the previous time step at time $t$. In other words, the displacement solution sought can be expressed as

$$\mathbf{U}^{t+\Delta t} = \mathbf{U}^t + \Delta \mathbf{U}^{\Delta t} \tag{37}$$

Due to nonlinearities, each increment has to be solved iteratively. Here, a Newton-Raphson iteration procedure will be introduced using the iteration counter $i$. Thus, the displacement increment is iteratively determined via

$$\mathbf{U}^{i+1} = \mathbf{U}^i + \Delta \mathbf{U}^{i+1} = \mathbf{U}^t + \sum_{k=1}^{i+1} \Delta \mathbf{U}^k \tag{38}$$

along with stresses and strains. For that purpose, a linearisation of Eq. (36) is performed around the current state given by the $i^{\text{th}}$ global Newton-Raphson iteration. For more details we refer the reader to Bathe (2001); Zienkiewicz et al (2006); Bucher et al (2001). In a Total Lagrangian setting, linearisation of Eq. (36) yields under the assumption of conservative loads

$$\int_{\Omega^0} \left[ \mathbf{S}^i : \bar{\mathbf{E}}(\Delta \mathbf{U}^{i+1}, \mathbf{V}) + \mathbf{E}(\mathbf{U}^i, \mathbf{V}) : \left.\frac{\mathrm{d}\mathbf{S}}{\mathrm{d}\mathbf{E}}\right|_i : \mathbf{E}(\mathbf{U}^i, \Delta \mathbf{U}^{i+1}) \right] \mathrm{d}\Omega^0 =$$
$$= \int_{\Omega^0} \rho_0 \mathbf{F}^{t+\Delta t} \cdot \mathbf{V}\, \mathrm{d}\Omega^0 + \int_{\partial \Omega_\mathbf{T}^0} \bar{\mathbf{T}}^{t+\Delta t} \cdot \mathbf{V}\, \mathrm{d}\Gamma^0 - \int_{\Omega^0} \mathbf{S}^i : \mathbf{E}(\mathbf{V}, \mathbf{U}^i)\, \mathrm{d}\Omega^0 \tag{39}$$

$$\mathbf{E}(\mathbf{U}^i, \mathbf{V}) = \text{sym}\left[\mathbf{F}^{i\mathrm{T}} \operatorname{Grad} \mathbf{V}\right] \quad (40)$$

$$\mathbf{E}(\mathbf{U}^i, \Delta\mathbf{U}) = \text{sym}\left[\mathbf{F}^{i\mathrm{T}} \operatorname{Grad} \Delta\mathbf{U}\right] \quad (41)$$

$$\bar{\mathbf{E}}(\Delta\mathbf{U}^{i+1}, \mathbf{V}) = \text{sym}\left[(\operatorname{Grad} \Delta\mathbf{U}^{i+1})^{\mathrm{T}} \operatorname{Grad} \mathbf{V}\right] \quad (42)$$

Eq. (39) introduced the constitutive tangent modulus in material description

$$\mathscr{C}_{\mathrm{m}} = \frac{\mathrm{d}\mathbf{S}}{\mathrm{d}\mathbf{E}} \quad (43)$$

Depending on the numerical scheme used (e.g., Total Lagrange, Updated Lagrange, or geometrically linear) different stress measures, strain measures and associated tangent moduli need to be considered. The choice regarding Voigt- or Kelvin-mapping remains open for all formulations.

4.3 Discretisation of the weak form

In the sequel, a switch to a matrix-vector notation representative of the finite element implementation will be highlighted by italic bold-face symbols. Where necessary, the matrices obtained using the standard Voigt mapping are compared to the ones obtained by Kelvin mapping. Where no distinction is mentioned, the implementation is unaffected by this choice.

The domain of interest is split into standard finite elements characterised by a set of nodal shape functions $N^a(\mathbf{x})$. The sought solution vector $\boldsymbol{U}$ in a point is approximated by

$$\boldsymbol{U} \approx \widetilde{\boldsymbol{U}} = \boldsymbol{N}\widehat{\boldsymbol{U}} \quad \text{with} \quad \boldsymbol{N} = \begin{pmatrix} N^1 \cdots N^{n_{\mathrm{n}}} & 0\cdots 0 & 0\cdots 0 \\ 0\cdots 0 & N^1 \cdots N^{n_{\mathrm{n}}} & 0\cdots 0 \\ 0\cdots 0 & 0\cdots 0 & N^1 \cdots N^{n_{\mathrm{n}}} \end{pmatrix} \quad (44)$$

where $\widehat{\boldsymbol{U}} = \left[\widehat{U}_1^1 \cdots \widehat{U}_1^{n_{\mathrm{n}}} \ \widehat{U}_2^1 \cdots \widehat{U}_2^{n_{\mathrm{n}}} \ \widehat{U}_3^1 \cdots \widehat{U}_3^{n_{\mathrm{n}}}\right]^{\mathrm{T}}$ is the nodal displacement vector and $\boldsymbol{N}$ is the element matrix of shape functions, and $n_{\mathrm{n}}$ is the number of nodes. In the isoparametric concept employed here, the position vector $\boldsymbol{X}$ and the test function $\boldsymbol{v}$ are approximated likewise.

The displacement gradient's coordinates are arranged into a nine-dimensional vector

$$\nabla \boldsymbol{U} = \left[\widehat{U}_{1,1} \cdots \widehat{U}_{1,3} \ \widehat{U}_{2,1} \cdots \widehat{U}_{2,3} \ \widehat{U}_{3,1} \cdots \widehat{U}_{3,3}\right]^{\mathrm{T}} \quad (45)$$

calculated based on the gradient matrix $\boldsymbol{G}$:

$$\nabla \boldsymbol{U} = \boldsymbol{G}\widehat{\boldsymbol{U}} \quad \text{with} \quad \boldsymbol{G} = \begin{pmatrix} N_{,1}^1 \cdots N_{,1}^{n_{\mathrm{n}}} & 0\cdots 0 & 0\cdots 0 \\ N_{,2}^1 \cdots N_{,2}^{n_{\mathrm{n}}} & 0\cdots 0 & 0\cdots 0 \\ N_{,3}^1 \cdots N_{,3}^{n_{\mathrm{n}}} & 0\cdots 0 & 0\cdots 0 \\ 0\cdots 0 & N_{,1}^1 \cdots N_{,1}^{n_{\mathrm{n}}} & 0\cdots 0 \\ 0\cdots 0 & N_{,2}^1 \cdots N_{,2}^{n_{\mathrm{n}}} & 0\cdots 0 \\ 0\cdots 0 & N_{,3}^1 \cdots N_{,3}^{n_{\mathrm{n}}} & 0\cdots 0 \\ 0\cdots 0 & 0\cdots 0 & N_{,1}^1 \cdots N_{,1}^{n_{\mathrm{n}}} \\ 0\cdots 0 & 0\cdots 0 & N_{,2}^1 \cdots N_{,2}^{n_{\mathrm{n}}} \\ 0\cdots 0 & 0\cdots 0 & N_{,3}^1 \cdots N_{,3}^{n_{\mathrm{n}}} \end{pmatrix} \quad (46)$$

The symmetric part of the displacement gradient corresponds to the linear part of the Green-Lagrange strain

$$\text{sym Grad } \mathbf{U} = \frac{1}{2} \left[ \text{Grad } \mathbf{U} + (\text{Grad } \mathbf{U})^\text{T} \right] =: \mathbf{E}_\text{lin} \tag{47}$$

Its mapping yields the linear $B_0$-Matrix familiar from small strain finite element implementations:

$$\mathbf{E}_\text{lin} = \mathbf{B}_0 \widehat{\mathbf{U}} \quad \text{with} \quad \mathbf{B}_0 = \begin{pmatrix} N^1_{,1} \ldots N^{n_\text{n}}_{,1} & 0 \cdots 0 & 0 \cdots 0 \\ 0 \cdots 0 & N^1_{,2} \ldots N^{n_\text{n}}_{,2} & 0 \cdots 0 \\ 0 \cdots 0 & 0 \cdots 0 & N^1_{,3} \ldots N^{n_\text{n}}_{,3} \\ a\left[N^1_{,2} \ldots N^{n_\text{n}}_{,2}\right] & a\left[N^1_{,1} \ldots N^{n_\text{n}}_{,1}\right] & 0 \cdots 0 \\ 0 \cdots 0 & a\left[N^1_{,3} \ldots N^{n_\text{n}}_{,3}\right] & a\left[N^1_{,2} \ldots N^{n_\text{n}}_{,2}\right] \\ a\left[N^1_{,3} \ldots N^{n_\text{n}}_{,3}\right] & 0 \cdots 0 & a\left[N^1_{,1} \ldots N^{n_\text{n}}_{,1}\right] \end{pmatrix} \tag{48}$$

In the case of Voigt mapping, $a = 1$ yields the familiar form. If Kelvin mapping is used, the linear $B_0$-matrix is obtained by setting $a = 1/\sqrt{2}$. In a small strain formulation, changing the value of $a$ in Eq. (48) remains the only change to be made (excluding input-output functions for pre- and postprocessing).

Nonlinear deformation measures of the form given in Eqs. (40) and (41) can be expressed as

$$\mathbf{E}(\mathbf{U}, \mathbf{A}) = \text{sym}\left[\mathbf{F}^\text{T} \text{Grad } \mathbf{A}\right] = \text{sym}\left[\text{Grad } \mathbf{A} + (\text{Grad } \mathbf{U})^\text{T} \text{Grad } \mathbf{A}\right] \tag{49}$$

and discretised using the nonlinear $B$-matrix as follows

$$\mathbf{E}(\mathbf{U}, \mathbf{A}) = \mathbf{B}\widehat{\mathbf{A}} = (\mathbf{B}_0 + \mathbf{B}_\text{N})\widehat{\mathbf{A}} \quad \text{with} \tag{50}$$

$$\mathbf{B}_\text{N} = \begin{pmatrix} U_{1,1}\left[N^1_{,1} \ldots N^{n_\text{n}}_{,1}\right] & U_{2,1}\left[N^1_{,1} \ldots N^{n_\text{n}}_{,1}\right] \\ U_{1,2}\left[N^1_{,2} \ldots N^{n_\text{n}}_{,2}\right] & U_{2,2}\left[N^1_{,2} \ldots N^{n_\text{n}}_{,2}\right] \\ U_{1,3}\left[N^1_{,3} \ldots N^{n_\text{n}}_{,3}\right] & U_{2,3}\left[N^1_{,3} \ldots N^{n_\text{n}}_{,3}\right] \\ a\left[U_{1,2}N^1_{,1} + U_{1,1}N^1_{,2} \ldots U_{1,2}N^{n_\text{n}}_{,1} + U_{1,1}N^{n_\text{n}}_{,2}\right] & a\left[U_{2,2}N^1_{,1} + U_{2,1}N^1_{,2} \ldots U_{2,2}N^{n_\text{n}}_{,1} + U_{2,1}N^{n_\text{n}}_{,2}\right] \\ a\left[U_{1,3}N^1_{,2} + U_{1,2}N^1_{,3} \ldots U_{1,3}N^{n_\text{n}}_{,2} + U_{1,2}N^{n_\text{n}}_{,3}\right] & a\left[U_{2,3}N^1_{,2} + U_{2,2}N^1_{,3} \ldots U_{2,3}N^{n_\text{n}}_{,2} + U_{2,2}N^{n_\text{n}}_{,3}\right] \\ a\left[U_{1,3}N^1_{,1} + U_{1,1}N^1_{,3} \ldots U_{1,3}N^{n_\text{n}}_{,1} + U_{1,1}N^{n_\text{n}}_{,3}\right] & a\left[U_{2,3}N^1_{,1} + U_{2,1}N^1_{,3} \ldots U_{2,3}N^{n_\text{n}}_{,1} + U_{2,1}N^{n_\text{n}}_{,3}\right] \end{pmatrix}$$

$$\tag{51}$$

$$\begin{pmatrix} U_{3,1}\left[N^1_{,1} \ldots N^{n_\text{n}}_{,1}\right] \\ U_{3,2}\left[N^1_{,2} \ldots N^{n_\text{n}}_{,2}\right] \\ U_{3,3}\left[N^1_{,3} \ldots N^{n_\text{n}}_{,3}\right] \\ a\left[U_{3,2}N^1_{,1} + U_{3,1}N^1_{,2} \ldots U_{3,2}N^{n_\text{n}}_{,1} + U_{3,1}N^{n_\text{n}}_{,2}\right] \\ a\left[U_{3,3}N^1_{,2} + U_{3,2}N^1_{,3} \ldots U_{3,3}N^{n_\text{n}}_{,2} + U_{3,2}N^{n_\text{n}}_{,3}\right] \\ a\left[U_{3,3}N^1_{,1} + U_{3,1}N^1_{,3} \ldots U_{3,3}N^{n_\text{n}}_{,1} + U_{3,1}N^{n_\text{n}}_{,3}\right] \end{pmatrix}$$

where $\mathbf{B}_\text{N}$ is the nonlinear part of the $B$-matrix. Identically to Eq. (48), $a = 1$ for Voigt mapping and $a = 1/\sqrt{2}$ for Kelvin mapping.

With the above definition, the linearised weak form in Eq. (39) can be discretised. Using the arbitrariness of the nodal values of the test function $\widehat{\mathbf{V}}$, one can write

$$\int_{\Omega^0} [\boldsymbol{G}^{\mathrm{T}} \bar{\boldsymbol{S}}^i \boldsymbol{G} + \boldsymbol{B}^{i\,\mathrm{T}} \boldsymbol{C}^i \boldsymbol{B}^i] \,\mathrm{d}\Omega^0 \, \Delta \hat{\boldsymbol{U}}^{i+1} = \int_{\Omega^0} \rho_0 \boldsymbol{N}^{\mathrm{T}} \boldsymbol{F}^{t+\Delta t} \,\mathrm{d}\Omega^0 + \int_{\partial \Omega_{\mathrm{T}}^0} \boldsymbol{N}^{\mathrm{T}} \bar{\boldsymbol{T}}^{t+\Delta t} \,\mathrm{d}\Gamma^0 - \int_{\Omega^0} \boldsymbol{B}^{i\,\mathrm{T}} \boldsymbol{S}^i \,\mathrm{d}\Omega^0 \qquad (52)$$

where $\bar{\boldsymbol{S}}$ is defined in C. The integral on the left hand side defines the stiffness matrix $\boldsymbol{K}$, the right-hand side defines the residual vector $\boldsymbol{\psi}$ such that the linearised system reads

$$\boldsymbol{K}^i \Delta \hat{\boldsymbol{U}}^{i+1} = \boldsymbol{\psi}^i \qquad (53)$$

The contributions of all elements are assembled into the global problem which is then solved for the vector of unknown displacement increments $\Delta \hat{\boldsymbol{U}}^{i+1}$.

It remains to be noted that in the case of Kelvin mapping

$$\boldsymbol{C} = \begin{pmatrix} C_{1111} & C_{1122} & C_{1133} & \sqrt{2} C_{1112} & \sqrt{2} C_{1123} & \sqrt{2} C_{1113} \\ C_{2211} & C_{2222} & C_{2233} & \sqrt{2} C_{2212} & \sqrt{2} C_{2223} & \sqrt{2} C_{2213} \\ C_{3311} & C_{3322} & C_{3333} & \sqrt{2} C_{3312} & \sqrt{2} C_{3323} & \sqrt{2} C_{3313} \\ \sqrt{2} C_{1211} & \sqrt{2} C_{1222} & \sqrt{2} C_{1233} & 2 C_{1212} & 2 C_{1223} & 2 C_{1213} \\ \sqrt{2} C_{2311} & \sqrt{2} C_{2322} & \sqrt{2} C_{2333} & 2 C_{2312} & 2 C_{2323} & 2 C_{2313} \\ \sqrt{2} C_{1311} & \sqrt{2} C_{1322} & \sqrt{2} C_{1333} & 2 C_{1312} & 2 C_{1323} & 2 C_{1313} \end{pmatrix} \qquad (54)$$

while in the case of Voigt mapping, $\boldsymbol{C}$ directly contains the entries $C_{ijkl}$ without any factors. The manipulation need not be performed manually in cases where the local stress update as described in the next section is consistently performed with Kelvin-mapped quantities.

4.4 Integration of constitutive models – local stress update algorithm

The equations necessary to integrate the stress increment usually lead to a differential-algebraic equation system (Hartmann et al 1997; Bucher et al 2001) of the form

$$\boldsymbol{0} = \boldsymbol{r}(\boldsymbol{z}, \boldsymbol{\epsilon}^i) \qquad (55)$$

where $\boldsymbol{r}$ represents the residual vector describing the evolution equations for stresses and internal variables, as well as constraints (e.g., the consistency condition in elasto-plasticity). Note, that in the local iterations to solve the above equation system, $\boldsymbol{\epsilon}^i$ from the global iteration is considered fixed and that a suitable time discretisation is assumed to have been performed already in the formulation of the residual vector. The state vector $\boldsymbol{z}$ contains the stress vector as well as all kinds of inelastic internal state variables:

$$\boldsymbol{z} = (\boldsymbol{\sigma}^{\mathrm{T}}, \boldsymbol{\kappa}_k^{\mathrm{T}}, \kappa_k)^{\mathrm{T}} \qquad (56)$$

Different methods exist to integrate inelastic constitutive models Zienkiewicz and Cormeau (1974); Doghri (1995); de Borst and Heeres (2002); Safaei et al (2015). Here, we consider a fully-implicit backward Euler scheme relying in a local Newton-Raphson iteration. A Taylor series expansion of the differential-algebraic system yields the iteration procedure for the local stress integration

$$-\boldsymbol{r}^j = \left.\frac{\partial \boldsymbol{r}}{\partial \boldsymbol{z}}\right|_j \Delta \boldsymbol{z}^{j+1} \qquad (57)$$

Once the iteration has converged, the use of the total differential of $r$ directly yields the consistent tangent matrix for the global iteration:

$$\frac{\mathrm{d}r}{\mathrm{d}\epsilon^{t+\Delta t}} = \frac{\partial r}{\partial \epsilon^{t+\Delta t}} + \left(\left.\frac{\partial r}{\partial z}\right|_{t+\Delta t}\right)\frac{\mathrm{d}z}{\mathrm{d}\epsilon^{t+\Delta t}} = \mathbf{0} \tag{58}$$

The first entry of the solution $\mathrm{d}z/\mathrm{d}\epsilon^{t+\Delta t}$ to the resulting linear system

$$\left(\left.\frac{\partial r}{\partial z}\right|_{t+\Delta t}\right)\frac{\mathrm{d}z}{\mathrm{d}\epsilon^{t+\Delta t}} = -\frac{\partial r}{\partial \epsilon^{t+\Delta t}} \tag{59}$$

is the sought tangent matrix $C^i$. This approach is not only beneficial for achieving the best possible convergence of the global problem due to an algorithmically consistent linearisation (Simo and Hughes 1998), but also yields $C^i$ consistent with Eq. (52) for both the Voigt and the Kelvin Mapping without any further modification.

## 5 Discussion

While the Voigt mapping is a perfectly viable option and is the basis of many consistent finite element implementations, the different treatment of the tensor-mathematical quantities and their matrix-mapped counterparts creates a lot of room for error and necessitates double implementations of the same mathematical concept for different types of quantities. In contrast, all the complications associated with the Voigt mapping can be avoided using the Kelvin mapping and come at little, if any, additional cost when using traditional finite element analysis software. Recently, $B$-matrix free implementations have been proposed Planas et al (2012) into which some of the considerations outlined above can be included as well. The motivation for the treatment outlined in Planas et al (2012) were related to difficulties caused by the Voigt mapping, many of which can be avoided with Kelvin's approach as well due to the formalised mapping indicated in Eqs. (17) and (18). Note further, that the present approach equally does not require an explicit calculation of the $D/C$-matrix, as it naturally follows from the integration of the material model outlined in Section 4.4.

In summary,

- Tensor norms are calculated identically without distinguishing stress or strain-type quantities.
- Tensor equations can be directly transformed into matrix equations without additional considerations.
- The only implementational changes are related to a scalar factor in the $B$-matrices for both large and small strain formulations and modified input/output functions for the use of the usual pre- and postprocessing tools.

**Acknowledgements**

The authors would gratefully like to acknowledge the funding provided by the German Ministry of Education and Research (BMBF) for the ANGUS+ project, grant number 03EK3022, as well as the support of the Project Management Jülich (PTJ). Additional funding was provided by the Helmholtz Initiating and Networking Fund through the NUMTHECHSTORE project.

## A Spherical and deviatoric projections

Any second order tensor $\mathbf{A}$ can be additively decomposed into a spherical and a deviatoric part:

$$\mathbf{A} = \mathbf{A}^S + \mathbf{A}^D \quad \text{with} \quad \mathbf{A}^S = \frac{1}{3}(\mathbf{A}:\mathbf{I})\mathbf{I} \tag{60}$$

where $\mathbf{I} = \mathbf{g}^k \otimes \mathbf{g}_k$ is the metric tensor formed by the contra- and covariant basis vectors, respectively. The mapping can also be written in terms of fourth order tensors:

$$\mathbf{A}^D = \mathscr{P}^D : \mathbf{A} \quad \text{and} \quad \mathbf{A}^S = \mathscr{P}^S : \mathbf{A} \tag{61}$$

with the fourth order tensors

$$\mathscr{P}^S = \frac{1}{3}\mathbf{I} \otimes \mathbf{I} \tag{62}$$

$$\mathscr{P}^D = (\mathbf{I} \otimes \mathbf{I})^{\overset{23}{T}} - \frac{1}{3}\mathbf{I} \otimes \mathbf{I} = \mathscr{I} - \mathscr{P}^S \tag{63}$$

Note in passing, that the deviatoric representation of a quantity $\underline{a}$ can be calculated with the projection

$$\underline{a}^D = \underline{\underline{P}}^D \underline{a} \quad \text{with} \quad \underline{\underline{P}}^D = \begin{pmatrix} \frac{2}{3} & -\frac{1}{3} & -\frac{1}{3} & 0 & 0 & 0 \\ -\frac{1}{3} & \frac{2}{3} & -\frac{1}{3} & 0 & 0 & 0 \\ -\frac{1}{3} & -\frac{1}{3} & \frac{2}{3} & 0 & 0 & 0 \\ 0 & 0 & 0 & 1 & 0 & 0 \\ 0 & 0 & 0 & 0 & 1 & 0 \\ 0 & 0 & 0 & 0 & 0 & 1 \end{pmatrix} \tag{64}$$

which is independent of the mapping used, i.e. $\underline{\underline{P}}^D = \boldsymbol{P}^D$ and thus $\boldsymbol{a}^D = \boldsymbol{P}^D \boldsymbol{a}$.

## B Notes on the pull-back of the weak form

The left-hand side of Eq. (35) is pulled back into the reference configuration by using $d\Omega = J d\Omega^0$ where $J = \det \mathbf{F}$ is the volume ratio. The pull-back proceeds as follows:

$$\int_{\Omega} \boldsymbol{\sigma} : \operatorname{grad} \mathbf{v} \, d\Omega = \int_{\Omega^0} J\boldsymbol{\sigma} : \operatorname{sym} \operatorname{grad} \mathbf{v} \, d\Omega^0 = \int_{\Omega^0} J\boldsymbol{\sigma} : \mathbf{F}^{-T} \underbrace{\frac{1}{2}\left[(\operatorname{Grad} \mathbf{v})^T \mathbf{F} + \mathbf{F}^T \operatorname{Grad} \mathbf{v}\right]}_{:= \mathbf{E}(\mathbf{v};\mathbf{u})} \mathbf{F}^{-1} \, d\Omega^0 \tag{65}$$

$$= \int_{\Omega^0} J \mathbf{F}^{-1} \boldsymbol{\sigma} \, \mathbf{F}^{-T} : \mathbf{E}(\mathbf{u};\mathbf{v}) \, d\Omega^0 = \int_{\Omega^0} \mathbf{S} : \mathbf{E}(\mathbf{u};\mathbf{v}) \, d\Omega^0 \tag{66}$$

where the second Piola-Kirchhoff stress $\mathbf{S}$ appears. The volume integral on the right-hand side of Eq. (35) is transformed similarly noticing that $\rho_0 = J\rho$ and that a vector can be expressed in both the reference and the current configuration by the use of shifters: $\mathbf{a} = a^i \mathbf{g}_i = a^i g_i^K \mathbf{G}_K = a^K \mathbf{G}_K$, where the coordinates of the shifter are $g_i^K = \mathbf{g}_i \cdot \mathbf{G}^K$. The surface traction $\mathbf{t} = \boldsymbol{\sigma} \mathbf{n}$ acting on the current (deforming) Neumann-boundary with (deformation dependent) area elements $d\Gamma$ can be transformed to the reference configuration with (constant) area elements $d\Gamma^0$ with the help of Nanson's formula:

$$\boldsymbol{\sigma} \mathbf{n} \, d\Gamma^{t+\Delta t} = \boldsymbol{\sigma} J \mathbf{F}^{-T} \mathbf{N} \, d\Gamma^0 = \mathbf{P} \mathbf{N} \, d\Gamma^0 = \bar{\mathbf{T}} \, d\Gamma^0 \tag{67}$$

Therefore, the pulled-back version of the linear momentum balance reads

$$\int_{\Omega^0} \mathbf{S} : \mathbf{E}(\mathbf{U};\mathbf{V}) \, d\Omega^0 = \int_{\Omega^0} \rho_0 \mathbf{F} \cdot \mathbf{V} \, d\Omega^0 + \int_{\partial \Omega_T^0} \bar{\mathbf{T}} \cdot \mathbf{V} \, d\Gamma^0 \tag{68}$$

where the vectors $\mathbf{v}$, $\mathbf{u}$ and $\mathbf{f}$ have been capitalised due to the fact that the basis of the reference configuration is chosen throughout.

## C Additional definitions

The stress matrix is defined as

$$\bar{\mathbf{S}} = \begin{pmatrix} \begin{matrix} S_{11} & S_{12} & S_{13} \\ S_{12} & S_{22} & S_{23} \\ S_{13} & S_{23} & S_{33} \end{matrix} & \mathbf{0} & \mathbf{0} \\ \mathbf{0} & \begin{matrix} S_{11} & S_{12} & S_{13} \\ S_{12} & S_{22} & S_{23} \\ S_{13} & S_{23} & S_{33} \end{matrix} & \mathbf{0} \\ \mathbf{0} & \mathbf{0} & \begin{matrix} S_{11} & S_{12} & S_{13} \\ S_{12} & S_{22} & S_{23} \\ S_{13} & S_{23} & S_{33} \end{matrix} \end{pmatrix} \quad (69)$$